\documentclass[aps,prl,amssymb,superscriptaddress,footinbib,twocolumn]{revtex4} 
\usepackage{graphicx}
\usepackage{amsmath}
\usepackage{color}

\begin{document}

\title{Vortices and turbulence in trapped atomic condensates}
\author{Angela C. White}
\email{ang.c.white@gmail.com}
\affiliation{Joint Quantum Centre (JQC), Durham-Newcastle, School of Mathematics and Statistics, Newcastle University, Newcastle upon Tyne, NE1 7RU, UK.}
\author{Brian P. Anderson}
\affiliation{College of Optical Sciences, University of Arizona, Tucson, AZ 85721, USA}
\author{Vanderlei S. Bagnato}
\affiliation{Instituto de F\'{\i}sica de S\~{a}o Carlos, Universidade de S\~{a}o Paulo, Caixa Postal 369, 13560-970 S\~{a}o Carlos, SP, Brazil.}


\begin{abstract}
After over a decade of experiments generating and studying the physics of quantized vortices in atomic gas Bose-Einstein condensates, research is beginning to focus on the roles of vortices in quantum turbulence, as well as other measures of quantum turbulence in atomic condensates.  Such research directions have the potential to uncover new insights into quantum turbulence, vortices and superfluidity, and also explore the similarities and differences between quantum and classical turbulence in entirely new settings. Here we present a critical assessment of theoretical and experimental studies in this emerging field of quantum turbulence in atomic condensates. 
\end{abstract}

\maketitle

\section{Introduction}

{S}ince Onsager's groundbreaking theoretical work linking turbulence and point vortex dynamics in a two-dimensional (2D) fluid \cite{Onsager1949}, it has been hoped that the simple nature of quantum vortices in superfluids will aid in understanding the nature of turbulence.  After many years of research with superfluid helium systems, the field of quantum turbulence (QT) is now well established, and has led to numerous new insights and developments regarding QT and the universality of turbulence \cite{QVDSF}.  The discovery of links between classical turbulence and QT remains a strong motivating factor for QT research, particularly in the emerging field of QT studies with Bose-Einstein condensates (BECs).  BECs present a new platform for QT studies due to their compressibility, weak interatomic interactions, and availability of new experimental methods for probing and studying superfluid flow  \cite{Pet2008}.  The relationship between QT and vortex dynamics in these systems is consequently an inherently interesting new research topic as well.

Classical turbulence is composed of eddies of continuous vorticity and size and it is necessary to solve the Navier-Stokes equation to mathematically describe viscous fluid dynamics \cite{Davidson2004}. For turbulent fluid flow, which consists of scale-invariant flow dynamics  across a wide range of length scales, this procedure becomes difficult to tackle from first principles.  In comparison, QT is comprised of vortices of less complexity, each with a localized and well-defined vortex core structure and quantized circulation. Superfluid flow is inviscid and vortices cannot decay by viscous diffusion of vorticity: a quantized vortex cannot simply ``spin down'' and dissipate energy via viscosity in the same way a classical vortex can.  Incompressible kinetic energy is instead diffused through emission of sound waves and then dissipated due to the presence of a thermal cloud in BECs or the normal fluid component in superfluid He. 

Despite the differences arising from the nature of vortices, classical and quantum turbulence share profound similarities that underscore the universality of turbulence.  We briefly illustrate this idea with the structure of kinetic energy spectra in three-dimensional (3D) turbulence.  At locations in the fluid far from vortex cores, and for length scales greater than the average inter-vortex spacing, vortex core structure is unimportant and quantized vortex lines are analogous to vortex filament lines of an Euler fluid.  Under these conditions,  classical and quantum turbulence are known to possess similar macroscopic and statistical properties. The most striking similarity is the existence of the same Kolmogorov spectrum in the inertial range for 3D turbulence.  For non-equilibrium steady-state forced turbulence, at length scales larger than the distance given by the average inter-vortex separation and smaller than the scale corresponding to energy injection, the incompressible kinetic energy spectrum scales with wavenumber $k$ as $E(k)\sim k^{-5/3}$ \cite{Maurer1998}.  This scaling is thought to arise from a Richardson cascade process as large vortices break up into smaller and smaller vortices until at very small scales energy is dissipated.  The same classical Kolmogorov scaling has been verified numerically and experimentally in turbulent superfluid $^{4}$He and $^{3}$He-B \cite{Maurer1998,Nore1997,Stalp1999,Araki2002,Bradleyqc,Salort2010PF,Baggaley2011PRE,Baggaley2011PRB} and has also been established numerically as a feature of trapped \cite{Kobayashi2007,Kobayashi2008} and homogeneous \cite{Kobayashi2005,Sasa2011} atomic condensates.  There is thus reason to believe that a quantitative understanding of aspects of turbulence in one system, even a quantum fluid, may aid in the general understanding of the subject.

Among the central issues in the development of an understanding of QT is therefore understanding how quantized vortices move about in a plane in 2D QT, or how they bend, interact, and create complex tangles in 3D QT. The relationship to vortex dynamics and statistical measures of turbulence, such as energy and velocity spectra, is also an ongoing and active research topic.  While the physics of such processes and the links between the various aspects of the problems may remain intricate, the reduction of vorticity to well-defined cores enables new approaches to studying QT generally unattainable in classical turbulence.  This article is aimed at presenting a brief overview of theoretical, numerical, and experimental progress --- as well as potential capabilities --- for achieving a deeper understanding of QT by tackling these central issues.  We present atomic BECs as a new and distinct tool that will contribute towards an extended understanding of QT and vortex dynamics through both theory and experiment.  Firstly we discuss the unique aspects of turbulence in BECs in comparison to superfluid $^{4}$He and $^{3}$He-B. We examine the recent theoretical and experimental advances in the BEC turbulence field and detail the new regimes of turbulence made accessible in BECs. Finally we overview experimental measures of turbulence and future directions for the field, focusing on progress in both 3D and 2D QT. 

\section{Atomic BECs: beyond superfluid helium}

Studies of quantum turbulence have historically been restricted to superfluid $^{4}$He and $^{3}$He-B. These systems have a huge range of accessible length scales and as a result, turbulent vortex tangles can consist of hundreds of thousands of vortices.  The vortex tangles are well separated with typical inter-vortex separation distances of  $l\sim10^{-4}$m. Vortices have small vortex core diameters, $\xi\sim10^{-10}$m for superfluid $^4$He, and consequently the turbulent tangles in  $^{4}$He and $^{3}$He-B are characterised by a large ratio of inter-vortex spacing to vortex core radius, on order of $10^{5}-10^{6}$.  The intrinsic superfluid parameters such as atom-atom interaction strength are fixed, and the superfluids are homogeneous and of constant density (i.e., incompressible). Controlling single-vortex dynamics in a turbulent $^{4}$He and $^{3}$He-B superfluid is exceptionally challenging, and probing the behaviour of turbulence at small scales remains an open problem from an experimental standpoint.  Finally, due to the strong interactions between atoms in liquid and superfluid helium, simulations of wave function dynamics are generally only qualitatively accurate.

In contrast, trapped atomic condensates are rarely homogeneous and have a smaller range of accessible length scales over which vortex dynamics can be probed. With the exception of shallow traps and nominally hard-wall confining potentials where the condensate is to a good approximation homogeneous, for realistic trapped systems the condensate density is non-uniform, with the non-uniformity arising as a consequence of the form of the trapping potential and the compressibility of BEC systems.  It is this compressibility that also lends significant new approaches to the study of QT.  Also because of this, manipulating the trapping potential allows one to manipulate the condensate homogeneity throughout the trapping region.  The typical vortex core diameter in atomic BECs is much larger than that in superfluid He,  being on order of the coherence or healing length, $\sim0.5\mu$m. Turbulent BECs consist of small numbers of vortices (generally no more than $\sim 100$)  that are less sparsely separated (ratio of inter-vortex spacing to vortex core radius on the order of 10).  From a theoretical standpoint, the models describing vortex tangles in superfluids are amenable to numerical simulation, and for BECs are well established and quantitatively accurate at both zero and non-zero temperatures \cite{FTProukakis,FTBlakie}.

BECs also permit adjustment of intrinsic atomic properties that lead to macroscopic flexibility not readily achieved in other superfluid systems. The strength of atom-atom interactions can be controlled and even driven from attractive to repulsive by tuning an external magnetic field around a Feshbach resonance \cite{RevModPhys.82.1225}.   Individual vortex dynamics and position can be well controlled \cite{DavisPRA2009,Samson2012}, and combined with a wide variety of imaging techniques opens up the field to investigations of few-vortex systems and the relationship of chaotic vortex dynamics to turbulent flow.  The tunable features of trapped atomic condensates further extends to their dimensionality.  Highly oblate condensates, with vortex dynamics well into the 2D regime can be routinely created. This makes BECs the first system for which 2D QT is experimentally accessible \cite{Samson2012,Neely2010phd,Neely2012,Wilson2013}.  We note that for such studies, quasi-2D BECs that have chemical potential less than the most strongly confining dimension's mode spacing are {\it not} needed.  Rather, highly oblate but nevertheless 3D BECs may suppress superfluid flow along the tight-confining direction enough that 2D superfluid vortex dynamics and 2D turbulence are obtained \cite{RooneyPRA2011}.  We will henceforth refer to such BECs as 2D, and the associated conditions, where vortex dynamics occur within a plane, as being 2D.  In addition to creating conditions for studying both 2D turbulence and 3D turbulence, BEC trapping parameters are extremely flexible, opening up the possibility of investigating transitions between 2D and 3D QT. 

\subsection{Theoretical formalism}

The dynamics of quantized vortices in zero-temperature BECs with scalar order parameter are described by evolution of the Gross-Pitaevskii equation,
\begin{equation}
i \hbar\frac{\partial \psi\left({\bf{r}},t\right)}{\partial t} = \left[-\frac{\hbar^{2}}{2m}\nabla^{2}+V\left({\bf{r}},t\right) + U_{0} |\psi\left({\bf{r}},t\right)|^{2}\right]\psi\left({\bf{r}},t\right)\,.
 \end{equation}
Here $\psi$ is a classical mean-field wave function representing the condensate trapped by a potential $V\left({\bf{r}},t\right)$. $U_{0}=4\pi\hbar^{2}a/m$ describes interactions between bosons of mass $m$ in the condensate, where $a$ is the $s$-wave scattering length, and $\hbar$ is the reduced Plank's constant \cite{Pet2008}. For simplicity, we consider BECs in cylindrically symmetric harmonic trapping potentials.  When the axial trapping frequency is much greater than the radial trapping frequency, $\omega_{z}\gg\omega_{r}$, the condensate will be highly oblate, enabling studies of 2D QT and vortex dynamics limited to the radial plane.  For $\omega_z \sim \omega_r$, 3D QT may be explored.  The dynamics of condensates at non-zero temperatures can be accurately described by applying classical field methods \cite{FTBlakie}. For a description of non-zero-temperature models that could be applied to superfluid turbulence,  we refer the interested reader to the reviews \cite{FTProukakis,FTBlakie}. 

\subsubsection{Compressible and Incompressible kinetic energy}

Decomposing the condensate kinetic energy into compressible and incompressible parts is a useful technique frequently applied to analyse how kinetic energy due to vortex lines and sound is distributed over length scales \cite{Nore1997}.  This is performed by defining a density weighted velocity field ${\bf{\Upsilon}}=\sqrt{n}\textbf{v}$, in terms of the superfluid velocity, $\textbf{v}=\frac{\hbar}{m}\nabla\theta$, where $n$ and $\theta$ are the position-dependent condensate density and phase profiles respectively.  Utilizing the fundamental theorem of vector calculus, we can write ${\bf{\Upsilon}} = {\bf{\Upsilon}}^{i}+{\bf{\Upsilon}}^{c}$, where ${\bf{\Upsilon}}^{i}$ is an incompressible (i.e., divergence-free) component satisfying $\nabla\cdot{\bf{\Upsilon}}^{i}=0$, and ${\bf{\Upsilon}}^{c}$ is a compressible (i.e., irrotational) component for which $\nabla\times{\bf{\Upsilon}}^{c}=0$. The compressible and incompressible kinetic energy spectra $E^{i,c}_{\mathrm{kin}}(k)$  are defined by 
\begin{equation}
E^{i,c}_{\mathrm{kin}}=\frac{1}{2}\int \text{d}\textbf{r}|{\bf{\Upsilon}}^{i,c}|^{2}=\int_{0}^{\infty}\text{d}kE^{i,c}_{\mathrm{kin}}(k),
\end{equation}
where $k=|\vec{k}|$ is a wavenumber and $E^{i}_{\mathrm{kin}}$ and $E^{c}_{\mathrm{kin}}$ are the total incompressible and compressible kinetic energies in the system (per unit mass).
The flow of incompressible kinetic energy across wavenumbers and its relationship to vortex dynamics is one of the central and critically important and challenging issues in classical turbulence as well as QT.  An understanding of spectra in 2D and 3D QT thus relates directly to the dynamics and distribution of quantized vortices.  Considerable effort has been recently concentrated on constructing the incompressible kinetic energy spectrum and also the angle-averaged momentum and velocity distribution in order to determine their scaling in relation to the distribution of vortices in 2D \cite{Nowak2011,Nowak2012,Schole2012,Bradley2011,Kusumura2012} and 3D QT in condensates \cite{Nowak2011}.

\subsubsection{Quantum Pressure Energy}

Contributions to the quantum pressure energy,  
\begin{equation}
E_{\mathrm{qp}}=\frac{1}{2}\int\text{d}\textbf{r}|\nabla\sqrt{n}|^{2}\,,
\end{equation}
arise only where the condensate density varies sharply, such as at vortex cores, dark solitons, and other density discontinuities.  Under conditions where solitons and other features of the compressible component are negligible or damped, this quantity may provide a useful theoretical measure of the total vortex number in 2D condensates. For 3D condensates, quantum pressure energy scales with the vortex line length. While evaluating incompressible, compressible and quantum pressure energies and spectra is useful from a theoretical point of view, it remains an open problem how to measure these individual components experimentally.   However, velocity correlations and the angle-averaged momentum spectrum are quantities potentially experimentally accessible in atomic BECs and we outline some measurement prospects later in the article.

\section{Vortex generation}

A plethora of experimental methods are available to induce vortices in BECs \cite{Anderson2010}. We highlight a non-exhaustive selection of techniques that allow the preparation of well defined initial states from which to investigate the evolution or decay of turbulence. Deterministic induction of vortices into the condensate at precisely defined positions can be achieved by the controlled methods of phase imprinting, where the phase profile of the condensate is engineered \cite{Ketterle2002,Shibayama2011}, or by the coherent transfer of orbital angular momentum to a condensate by a two-photon stimulated raman process \cite{Andersen2006,Wright2008,Wright2009,Leslie2009}. The creation of non-equilibrium vortex states with arbitrary winding has been beautifully demonstrated by the transfer of orbital angular momentum from a holographically produced light beam \cite{Brachmann2011}. Applying this technique has the potential to create arbitrarily complex initial vortex distributions and even vortex knots in BECs \cite{Proment2012,Dennis2010}. Alternatively, vortex knots could be imprinted in BECs applying the techniques demonstrated in classical fluids, through the acceleration of shaped hydrofoils \cite{Kleckner2013}, made by a 3D laser structure or a shaped nano-tube in a condensate. 

Laser stirring can also be applied to create vortices in atomic BECs in a deterministic manner \cite{Neely2010}. The distribution of vortices throughout a condensate is known to depend on the path of the laser stirrer \cite{White2012}.  For a 2D condensate, this means the path the laser stirrer takes through the condensate can create clusters of like-signed vortices as shown in Fig.~1(A), or a more random distribution of vortices of differing sign.  For a 3D condensate, this implies that a stirring path might be optimised to generate well-distributed vortex configurations or more polarised tangles, where the vortex distribution is aligned preferentially along a particular direction.  Reeves et~al.\cite{Reeves2012} have shown that the potential strength of a laser stirrer and its speed can also be chosen such that the laser stirrer sheds single dipoles, clusters, or oblique solitons in a trapped 2D BEC.  These ideas should also extend to stirring or flow past an obstacle in a 3D condensate. 

In addition to laser stirring, combined rotation and precession around three cartesian axes has been shown to create isotropic vortex tangles in 3D condensates \cite{Kobayashi2008}. Reducing the rotation and precession of the condensate to only two directions decreases the degree of isotropy of the resulting vortex tangle and could also be applied to create polarised distributions of vortices \cite{Kobayashi2007}. The methods of laser stirring, phase imprinting, transfer of orbital angular momentum and combined rotation and precession which have been highlighted here, have the additional advantage of producing only small amounts of phononic excitations, where the acoustic energy density is much less than the incompressible energy density; see \cite{Reeves2012} for further discussion of this relationship.
 
\section{Chaos and few-vortex dynamics}

The ability to create well-defined initial distributions of vortices \cite{Samson2012} opens up the possibility of directly probing systems of few vortices and their resulting dynamics.  Previous experiments have observed the precession of single filled vortices in trapped condensates \cite{And2000.PRL85.2857}, and the motion of vortex dipoles deterministically created prior to expansion of a sequence of BECs \cite{Neely2010}.  Recent experimental advances have demonstrated measurements of few-vortex dynamics within a single BEC \cite{Freilich2010,Navarro2013}.  This method is accomplished by allowing only a small fraction of the atoms in a BEC to be imaged after expansion, and sequential images from multiple expansion steps allows determination of vortex dynamics.  This technique could be incorporated into future measurements of chaotic vortex dynamics.  With the future development of real-time in situ imaging of vortex dynamics it may become possible to directly observe vortex dynamics, including vortex-vortex annihilation and reconnection events directly within BECs.  The motion of four point vortices in a plane can be chaotic \cite{Aref1980,Aref1982,Aref1988} and similar chaotic dynamics are expected to be observable in oblate trapped BECs. The number of vortices that determines the crossover from chaotic to turbulent vortex dynamics and the role of chaos in turbulence \cite{Are1983.ARFM15.345} remain open questions that atomic BECs may be able to address.  

\section{Three-dimensional turbulence}

\subsection{Non-equilibrium steady-state turbulence}

One of the defining features of continuously forced turbulence in a bulk 3D fluid is the existence of a direct Kolmogorov cascade corresponding to the conserved flow of energy from the forcing scale to smaller length scales.  Although the direct Kolmogorov energy cascade has been established numerically for trapped atomic condensates \cite{Kobayashi2007,Kobayashi2008}, it is yet to be confirmed experimentally.  Indeed for a realistic tangle of vortices, where the BEC diameter is typically less than $\sim100\mu$m, the small range of length scales present may prohibit its detection. On the other hand, this implies trapped condensates may be a suitable system in which to determine the lower bound on the system size and vortex line density for which Kolmogorov scaling may be observed. Although the $k^{-5/3}$ scaling of the incompressible energy spectrum is established in superfluid He systems and is thought to arise due to a Richardson cascade process \cite{Maurer1998,Nore1997,Stalp1999,Araki2002,Salort2010PF,Baggaley2011PRE,Baggaley2012PRL}, a corresponding break up of large eddies into smaller and smaller eddies has not been theoretically or experimentally verified. If a Richardson cascade process is present in QT, it may be directly observable in trapped atomic condensates as detection schemes that experimentally image vortices directly are feasible.  We outline such techniques near the end of this article.  Tracking the length of individual vortices is also possible numerically, and may help answer questions about the relationship of an energy cascade to a Richardson cascade.  Establishing the existence or absence of a Richardson cascade process coupled to a direct Kolmogorov cascade is perhaps one of the most exciting prospects in future experimental and numerical studies of turbulent vortex tangles in trapped atomic BECs.

\subsection{Small-scale dynamics and dissipation}

\hspace{0.1mm} On length scales smaller than the typical inter-vortex spacing, quantum turbulence is quite different from classical turbulence. Characteristics that are dependent on the vortex core structure and microscopic nature of the fluid are expected to be unique for quantum fluids. One such property is the velocity statistics of turbulent quantum fluids.  The velocity components of trapped and homogeneous turbulent BECs in 2D and 3D, as well as turbulent superfluid He, follow power-law like behaviour \cite{Salort2010PF,Paoletti2008PRL,White2010,Adachi2011}, in stark contrast to gaussian velocity statistics of turbulent classical fluids \cite{Vincent1991,Noullez1997,Gotoh2002}. This difference in the velocity statistics of quantum fluids arises from the singular nature of quantized vorticity and the $1/r$ velocity field of a quantum vortex \cite{White2010,Min1996}. Measurement techniques in superfluid ${}^{4}$He can also probe the `quasi-classical' limit, that is they do not measure length-scales smaller than the average vortex separation distance. In this limit, velocity statistics lead to  a gaussian distribution \cite{Baggaley2011PRE,Salort2012}.  This is another demonstration of features of classical turbulence emerging for quantum turbulent systems, at large scales, where the microscopic vortex core structure is not probed.  

In 3D turbulence, the direct Kolmogorov cascade is coupled to dissipation of incompressible kinetic energy at small scales.  In superfluid He, at zero temperature this sink of incompressible kinetic energy is facilitated by a Kelvin wave cascade process at scales smaller than the average inter-vortex separation distance. The Kelvin-wave cascade process occurs when the helical perturbations on vortex lines cascade to shorter and shorter wavelengths, until at small scales phonons are excited, dissipating the kinetic energy of vortices \cite{VinenPRL2003}. For zero temperature condensates, such a Kelvin wave cascade process could also provide a sink of incompressible kinetic energy, in particular as Kelvin waves can be generated by vortex reconnection events \cite{KivotidesPRL} that are more frequent due to the smaller inter-vortex spacing in 3D condensates. As the collision of finite amplitude sound waves, or rarefaction waves with vortex lines can also generate Kelvin waves \cite{BerloffPRA2004}, they may feature prominently in tangles generated by methods that induce a significant quantity of sound.  Whether the Kelvin wave cascade process is driven by local or non-local transfer of energy remains an open and contentious question \cite{KozikPRL2004,LauriePRB2010,LebedevJLTP2010,KozikJLTP2010,LebedevJTLTPreply,LNLTP10,BouePRB2011,Sonin2012}. However, Gross-Pitaevskii equation simulations for Kelvin waves in a homogeneous system find scaling consistent with that predicted for a non-local energy cascade caused by 4-wave interactions \cite{KrstulovicPRE2012}.  Numerical simulations of few-vortex systems in 3D trapped condensates at non-zero temperature have shown Kelvin waves to be important in facilitating the decay of vortices by their movement out of the trap \cite{RooneyPRA2011}. This work also  showed that Kelvin mode excitations can be effectively frozen out by altering the condensate dimensionality through flattening the condensate, or increasing its confinement along the vortex line direction \cite{RooneyPRA2011}.  

In addition to the dissipation of incompressible kinetic energy through a Kelvin wave cascade process, for finite size trapped condensates, the dissipation of incompressible kinetic energy by sound generated through reconnection events \cite{Leadbeater01,Zuccher12} will also be important.  It has been suggested that the break up of vortex loops into smaller vortex rings could form a self-generating reconnection process to smaller and smaller scales until they are dissipated by self-annihilation and sound waves \cite{BayerPRB2011,Simula2011}. Whether sound produced by a Kelvin wave cascade process, or from reconnection events, is the dominant process dissipating incompressible kinetic energy in zero temperature trapped atomic condensates, and how this scales with condensate size, vortex line density, and the polarity of vortex tangles, remain open questions to be addressed.

\subsection{Decay of turbulence}

In addition to forced 3D turbulence, it is also possible to investigate decaying QT in trapped 3D BECs. The decay of turbulence is expected to be a complex process facilitated by vortex-vortex reconnections with the rate of decay of vortex line density also influenced by the condensate homogeneity and temperature.  A numerical study of the decay of a vortex tangle induced from straight vortices imprinted along each cartesian direction employed a phenomenological model of temperature and found vortex-line length to decay faster with increased dissipation, qualitatively modelling higher temperatures \cite{White2010}. No distinction between scalings of vortex line-length decay predicted from `Vinen' or `ultra-quantum' models of turbulence or `quasi-classical' models of turbulence describing decaying turbulence in superfluid He was observed. `Ultra-quantum' models of turbulence describe random distributions of vortices in superfluid He turbulence, with vortex decay facilitated by a Kelvin-wave cascade process corresponding to a decay of vortex line-density $L$ as $L\sim t^{-1}$ \cite{Golovqcvsuq,Baggaleyqcvsuq}. For turbulence in superfluid He where a Kolmogorov cascade features, vortices are locally aligned \cite{Volovik03}, and the decay of line density is observed to scale as $L\sim t^{-3/2}$ \cite{Bradleyqc,Golovqcvsuq,Baggaleyqcvsuq}.
A unique scaling of the decay of vortex line-length, dictated by the length-scale at which turbulence was forced, may feature in the decay of  turbulent tangles in trapped condensates. As the number of vortices in a turbulent BEC is much less than in turbulent superfluid He, yet vortex tangles are typically more dense, the processes dictating vortex decay are expected to be strongly influenced by the density of vortices and trapping inhomogeneity. 

\begin{center}
\begin{figure}
\includegraphics[width=0.98\linewidth]{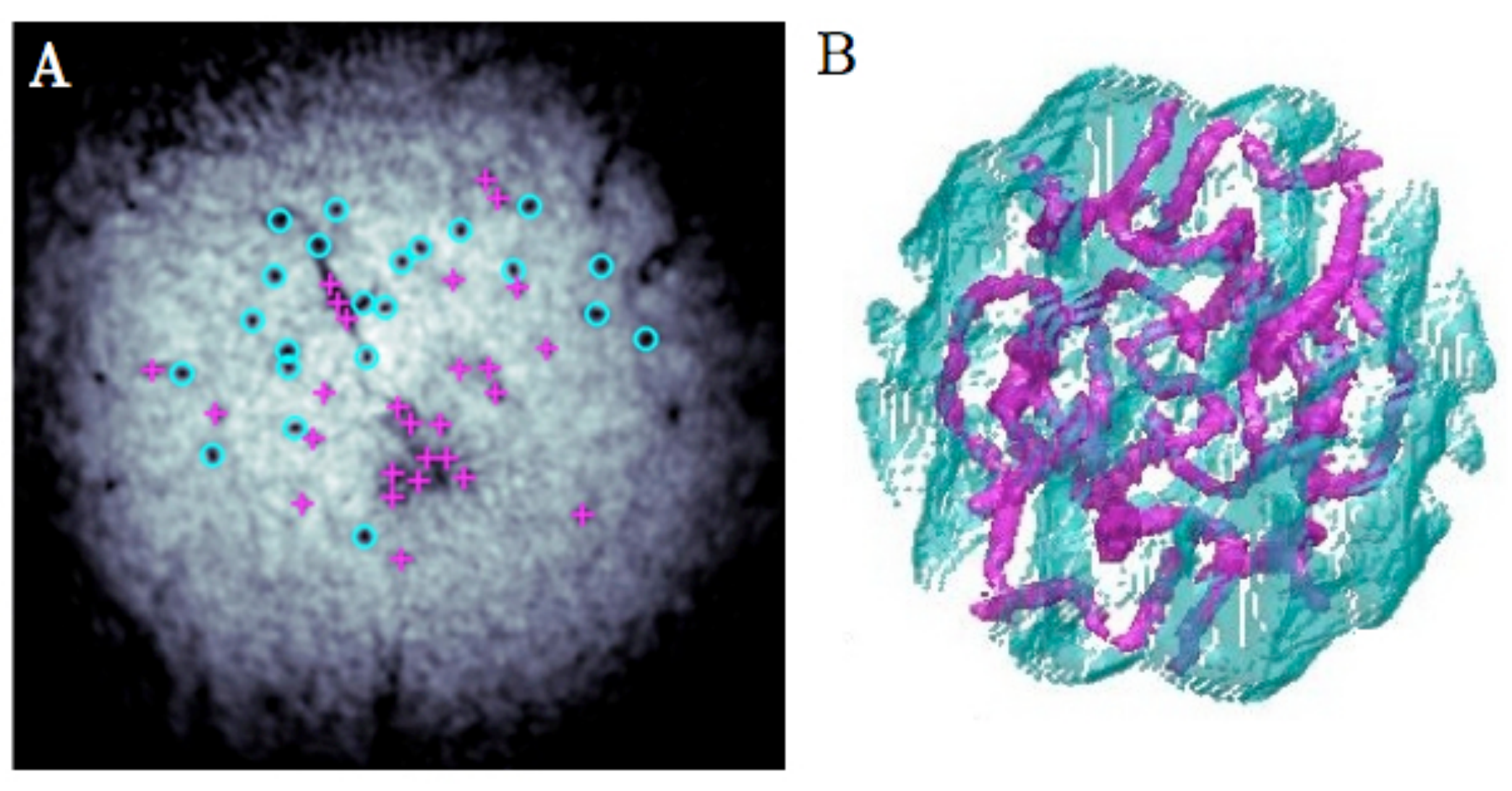}
\caption{A). Density profile of a 2D condensate where a cluster of like-signed vortices has been induced by stirring with a laser paddle as in \cite{White2012}. Vortices of positive and negative winding are denoted by magenta $+$ and cyan $-$ symbols respectively. B). 3D decaying turbulence: distribution of vortex lines (magenta) introduced by phase imprinting a lattice of $17$ straight vortices in a harmonically trapped condensate \cite{White2010}. The condensate edge defined by a density that is 25\% of the peak density is depicted in cyan. }
\end{figure}
\end{center}

\subsection{Experiments on three-dimensional QT}

\hspace{0.1mm} Between the years 1999 and 2009, numerous aspects of the physics and characteristics of quantized vortices were studied in experiments with superfluid atomic gases \cite{Anderson2010}.  These investigations included methods of vortex generation, imaging, and manipulation; properties of highly rotating condensates; vortex interactions, structure, stability; and relationships to superfluidity.  However, few references to turbulence can be found in this body of work, and appear primarily as brief qualitative statements given to refer to disordered distributions of vortices.  Such observations were made in the contexts of rotating single-component \cite{Che2000.PRL85.2223,Ram2001.PRL87.210402} and two-component \cite{Sch2004.PRL93.210403} BECs, and with fast laser sweeps through a BEC \cite{Ino2001.PRL87.080402}.  

In arguably the first experiment principally aimed at studying a turbulent tangle of
vortices in a 3D BEC, a non-uniform distribution of vortices was
created in a cigar-shaped condensate by exciting a surface mode
instability induced by an applied external oscillatory potential
\cite{Henn09a,Hennpra09}. Varying the strength and duration of the perturbing potential increased the number of vortices nucleated, enabling the observation of few-vortex configurations \cite{threevortex} through to a tangled distribution of vortices. In the regime of many vortices, it is difficult to identify individual vortex cores, particularly with detection of an expanded BEC's column density distribution. In these experiments on 3D QT, the turbulence was in a regime with few enough vortices that absorption images of expanded condensates showed the presence of vortex lines of varying orientations.  A tangled vortex configuration was then inferred from the absorption images. On release of the trap, the
condensate aspect ratio (defined by the measured ratio of BEC width to height) 
was conserved and self-similar expansion was
observed \cite{Henn09a}.  This behaviour is in stark contrast to the typical
inversion of the aspect ratio observed on expansion of a vortex-free
cigar shaped condensate and is thought to be an indication of a tangle
of vorticity throughout the condensate.  This conclusion is supported by theoretical investigations that studied the influence on the expansion of a cigar-shaped cloud arising from the presence of randomly distributed vorticity \cite{Caracanhas2013} and distributions of vorticity with preferred directions in a hydrodynamic framework \cite{Caracanhas2012}.  However, full numerical simulations going beyond the hydrodynamic limit and including quantum effects are still required. 
Measuring the condensate momentum distribution could potentially also provide useful information about the vortex distribution throughout the condensate \cite{Nowak2011}. Measurements towards this aim are underway at the Unversidade de S\~{a}o Paulo. 

\section{Two-dimensional turbulence}

The existence of the direct Kolmogorov cascade for 3D superfluid turbulence suggests that large scale features of 2D turbulent classical flow \cite{Kraichnan1980,Kellay2002,Bofetta2012} may also crossover to 2D turbulence in quantum fluids.  One of the characteristic features of classical 2D turbulence is the possible appearance of an inverse energy cascade (IEC) in which energy flow across length scales may occur in a direction opposite that of 3D turbulence; that is, with kinetic energy injected at small length scales, energy flows towards larger length scales in an inertial range free of forcing and dissipation mechanisms \cite{Kra1967.PF10.1417}.  This inertial range,  with $E_{\mathrm{kin}}^i(k) \propto k^{-5/3}$, corresponds to merging and growth of vortices; the IEC and vortex growth can continue until the largest wavelength modes of the system contain significant energy in the form of a vortex (or vortex dipole) on the scale of the size of the system.  The IEC is enabled by the 2D nature of the system, where local vorticity must be normal to the 2D plane.  This dimensional restriction gives rise to a conservation law for enstrophy (net squared vorticity) not present in 3D turbulence.  As conserved energy and enstrophy flux cannot simultaneously occur in two dimensions, 2D turbulence can simultaneously display a flux of energy towards length scales larger than that of forcing (the IEC), and flux of enstrophy towards smaller length scales (a direct enstrophy cascade).   The enstrophy cascade exhibits $E_{\mathrm{kin}}^i(k) \propto k^{-3}$ scaling over wavenumbers larger than that of the forcing scale.

Whereas the classical IEC of a 2D fluid is associated with the growth of patches of vorticity due to vortex merging and pushing energy towards larger length scales, the enstrophy cascade is associated with the stretching of patches of vorticity.   Experiments on forced and decaying 2D turbulence have shown that an IEC that corresponds to vortex merging can appear with an enstrophy cascade \cite{Rutgers1998a}, or may be observed without a simultaneous enstrophy cascade \cite{Som1986.JFM170.139,Par1997.PRL79.4162}.  Similarly, evidence for enstrophy cascades with  approximate $k^{-3}$ dependence in the energy spectrum, and corresponding observations of vortex stretching, may appear without an IEC particularly in decaying 2D turbulence \cite{PhysRevLett.74.3975,Mar1998.PRL80.3964}.  The type of forcing used in experiments may thus have a significant effect on the characteristics of observed spectra and flow dynamics.

These cascades are only beginning to be explored in 2D QT.  Among the most significant research topics within this field are the determination of  (i) conditions under which an IEC and an enstrophy cascade can be found in forced or decaying 2D QT, either together or individually; (ii) superfluid dynamics that accompany either cascade process, in particular the clustering of quantized vortices accompanying an IEC, which is the process most readily envisioned to correspond to vortex growth in 2D QT, and the appearance of large-scale flow that may be produced by energy flux into large scales; and (iii) vortex dynamics that may provide insight into the mechanisms underlying turbulent dynamics, such as vortex-antivortex annihilation and few-vortex dynamics.  The remainder of this section is devoted to addressing aspects of these open problems. 

In 2D QT enstrophy is proportional to the number of vortices \cite{Bradley2011,Numasato2010}.  In decaying 2D QT, enstrophy is rarely conserved throughout the entire system, as vortex-antivortex annihilation events occur \cite{Numasato2010,Numasato2010jltp} and serve as a dissipation mechanism. However, annihilation does not automatically preclude the appearance of an IEC in a forced system, and evidence for IECs have been observed in numerical simulations.  These will be discussed below. For decaying 2D QT, existence of IECs remains an open question. To complicate matters further, in smaller systems, vortex dynamics are strongly influenced by the condensate confining geometry.  For harmonically trapped BECs, the energy of a single vortex can be lost to the thermal cloud as the vortex precesses on a radial trajectory from the condensate centre towards the BEC boundary \cite{Freilich2010,Rosenbusch2002}.
 
Another effect potentially prohibitive to establishing IECs in BECs is the generation of large amounts of phononic excitations. Sound can be generated from vortex-dipole recombination events, the movement of vortices, and in particular by the method applied to induce vortices into the condensates.  Sound may also increase the frequency of vortex-antivortex annihilation events.  Furthermore, in consideration of Onsager's arguments \cite{Onsager1949}, there may be a minimum local number density of vortices or a forcing rate necessary to generate an IEC process.  These issues imply the presence of IEC processes in atomic BECs is no more clear cut than it is in classical turbulence, and considerable theoretical and experimental studies have been concentrated on determining conditions for its existence. It is furthermore necessary to establish the dominant parameters that dictate different regimes of vortex dynamics for continuously forced and decaying 2D QT, so that these various regimes can be associated with features of energy spectra, forcing, and dissipation.

To date, most of the work on compressible 2D QT has been numerical in nature.  We turn our attention to summarizing some of the findings that are among the most relevant for understanding vortex dynamics and energy cascades.  Decaying homogeneous 2D QT generated by random phase initial conditions was observed to exhibit a direct energy cascade with $k^{-5/3}$ scaling \cite{Numasato2010}.  In a study of vortex clustering, statistical measures of clustering were applied to quantify the distribution of vortices nucleated from a moving laser stirrer in a harmonically trapped condensate.  This study found no increase in clustering of vortices and only a significant degree of clustering was observed when the clustering was forced by the path of the laser stirrer \cite{White2012}.  In a different study using an annular trap, signs of weak clustering were observed, along with energy spectra consistent with (but not directly confirming) an IEC; this study was made in conjunction with experimental observations, and will be discussed at the end of this section \cite{Neely2012}.
Reeves et~al.\cite{Reeves2012} looked at a laser stirrer nucleating turbulent vortices 
and exciting acoustic energy into the BEC in order to distinguish regimes of vortex and acoustic turbulence.  It was found that in a particular regime of stirring where clustering of vortices of the same sign occurred,  Kolmogorov $k^{-5/3}$ scaling was observed, but became more transient as the fraction of vortices that were clustered decreased.  

Another recent 2D QT study of a homogeneous BEC simulated superfluid flow past past four static stirrers, and varied the degree of damping in the system.  Regimes of vortex clustering and energy spectra were then determined.  For one range of damping, incompressible kinetic energy spectra approximately proportional to $k^{-5/3}$ for length scales larger than that of the forcing scale were found \cite{Reeves2013PRL}.  Along with this spectral power law, a flux of incompressible kinetic energy into the longest wavelength modes of the system was measured.  The growth of vortex clusters was simultaneously observed: localized regions of vortices of the same sign were observed to grow in circulation quanta, and in size by a factor of approximately 5, both in time and distance away from the forcing region.  Together, these three observations show that IECs can be supported in 2D QT and that there is a correspondence between an IEC and the growth of patches of vorticity coinciding with vortex clusters, as in 2D classical turbulence.  Nevertheless, there still remain numerous open issues regarding vortex clustering and IECs that must be resolved.  These include a broad discovery of forcing and dissipation parameters that lead to the development of an IEC and vortex clustering, the roles and effects of boundary conditions and system geometry, the possibility of an enstrophy cascade developing, and the conditions under which a large region of vorticity on the scale of the system size might develop. 
 
Finally, an analytical approach to understanding the spectra of 2D QT for a compressible superfluid was pursued in \cite{Bradley2011}.  In this work, the shape of vortex cores in a BEC was used to analytically derive an expression for incompressible kinetic energy spectra of various vortex configurations in a homogeneous BEC.  Intriguingly, for small enough length scales, the spectrum is proportional to $k^{-3}$, although does this not correspond to an enstrophy cascade.   A particular type of vortex clustering was shown to lead to a $k^{-5/3}$ distribution in the spectrum for large enough length scales.   This paper proposed a scenario of forced 2D QT in which an IEC might appear over length scales larger than an assumed forcing scale at the lower end of the $k^{-3}$ range, although there is no enstrophy cascade in the proposed picture.  By considering the fraction of vortices involved in clusters, numerical results of forced turbulence correspond surprisingly well to the analytical predictions \cite{Reeves2013PRL}.  An analytical derivation of the Kolmogorov constant for forced 2D QT in a BEC was also proposed, suggesting that new approaches to 2D QT could eventually lead to new insights on 2D turbulence.
 
From the body of recent theoretical work performed to date, we can conclude that the IEC, vortex clustering, and system-scale vortex growth may appear in 2D QT, but understanding the full range of conditions for the appearance of these features will require much further exploration.  Simulations have shown that the manner in which vortices, as well as sound, are forced into the BEC can play a large role in the spectra observed, and this is one of the more complex aspects of the 2D QT field.   In the above summaries, we have focused on investigations that center on IECs and inertial regimes in energy spectra, but there are indeed other important aspects of 2D QT in BECs.   In particular, discussions of non-thermal fixed points lead to new means of understanding and discussing quantum turbulence \cite{Nowak2011,Nowak2012,Schole2012}, and studies of vortex dynamics and correlations may provide further means of analyzing regimes of QT \cite{Kusumura2012}.     
 
\subsection{Experiments on two-dimensional QT}

\begin{center}
\begin{figure}
\includegraphics[width=0.98\linewidth]{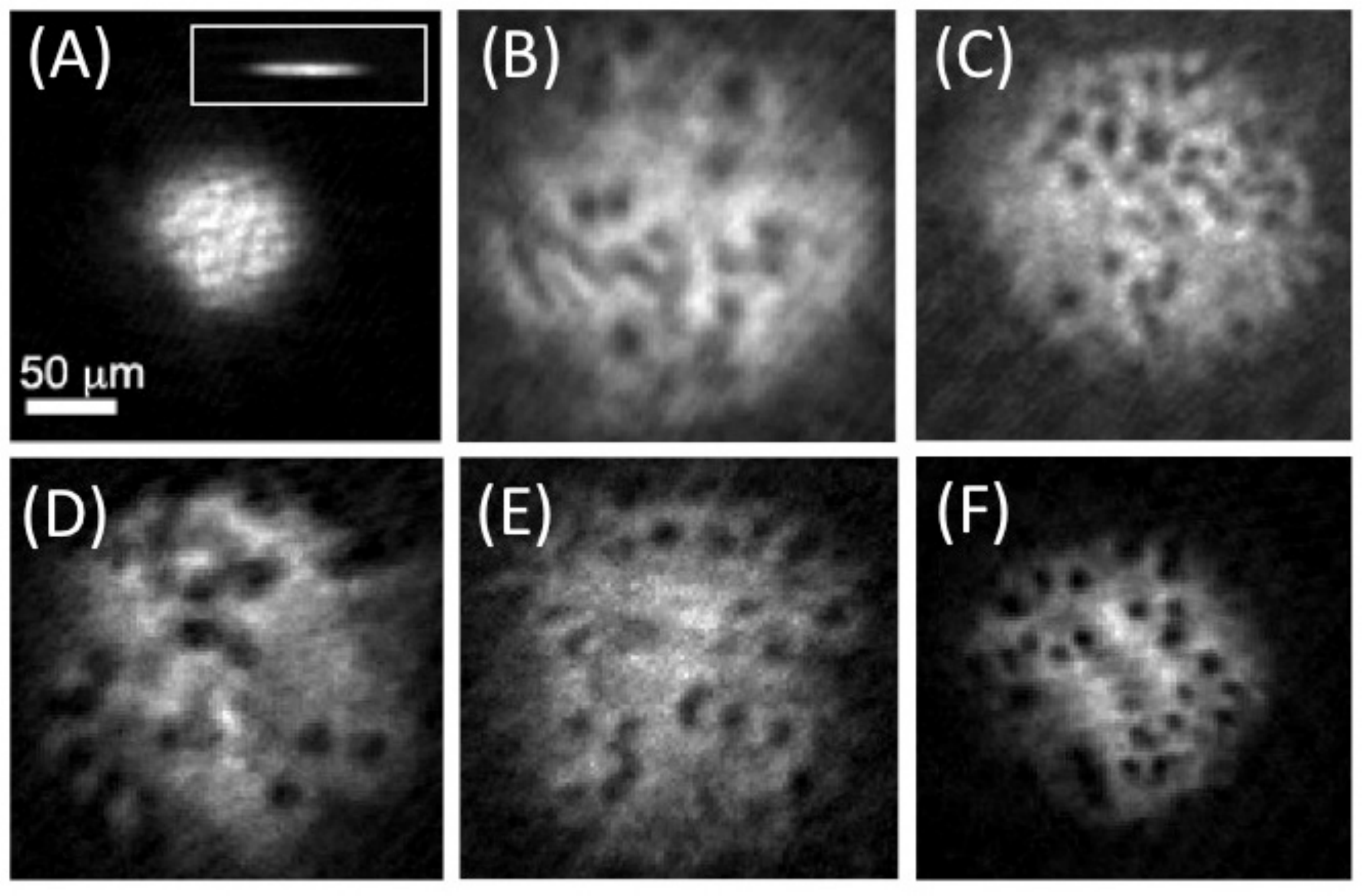}
\caption{\label{fig2DQTexpt}(A).  2D QT can be studied in a highly oblate BEC, as shown in an in-situ image of the BEC taken along the tightly confining axis and a corresponding in-situ image taken along the radial plane (inset). (B) In an annular BEC (created by directing a repulsive focused laser beam through the BEC along the tight trapping direction) excited by small-scale stirring using a repulsive laser potential, vortices were observed after the repulsive central laser barrier was removed prior to BEC expansion.   Note that the circulation of vortices can not be determined in this expansion method.  Taken from the experiment of \cite{Neely2012}. The remaining images of expanded BECs show 2D QT generation due to (C) modulation of the harmonic potential in a trap with a central focused repulsive laser potential, followed by 50 ms of hold time, (D) laser beam modulation (image taken 100 ms after removing beam), (E) sudden localized repulsive laser beam application and removal followed by 125 ms of hold time, and (F) trap spinning followed by 2.5 s of hold time.  In all cases, the BEC is expanded prior to imaging. See Ref.~\cite{Wilson2013} for details on these images and experimental methods.}
\end{figure}
\end{center}

\hspace{0.1mm}  As with 3D QT in BECs, there are still few experiments on 2D QT in trapped atomic condensates, and there is much to learn. The first 2D QT experiment by Neely et~al.\cite{Neely2012} observed the formation and evolution of disordered distributions of vortices forced into a highly oblate annular BEC by stirring with a laser beam, as shown in Fig.~\ref{fig2DQTexpt}(B). The technique excited very little acoustic energy, and injected on the order of 20 vortices, which may well be enough to display the physics of turbulence in such a small system.  Subsequent to stirring, the system evolved into a persistent current as vortex number decayed; while this observation corresponds to the growth of energy in large-scale flow given small-scale forcing, the extent to which this observation directly relates to the vortex and spectral dynamics described above is currently unclear.  Numerical simulations of the process showed the presence of pairs of like-signed vortices, an incompressible energy spectrum proportional to $k^{-5/3}$ over a range of length scales larger than the forcing scale, and conservation of enstrophy over the time period associated with the growth of the $k^{-5/3}$  spectrum.  While these observations are consistent with an IEC, the energy flux has not been determined, and the definite existence of an IEC in these simulations remains an open issue. 

Other methods have also been experimentally explored for producing 2D QT in highly oblate BECs \cite{Wilson2013}, as shown in Fig.~\ref{fig2DQTexpt}(C)-(F).  These include modulating the strength of the trapping potential, suddenly applying then removing a repulsive laser potential to a localized region in the BEC, modulating the intensity of a localized repulsive laser potential, and spinning a slightly elliptical highly oblate trapping potential within the radial plane.  Within the parameter ranges explored, all of these methods have been experimentally found to induce large numbers of vortices in harmonic and annular traps, and are candidates for further 2D QT studies.  Unfortunately for the study of vortex turbulence, these methods also generally excite acoustic or breathing modes of the BEC, and may render 2D QT studies difficult without further modification.  A deeper understanding the physical origins of vortex generation in each case is also needed.

\section{Quantum Turbulence in BECs: future prospects}

\subsection{Central experimental challenges}

Here we outline some of the main experimental challenges that must be overcome to explore aspects of  QT in BECs.  Ideally, an experiment would be able to watch all vortices in real time, examine their dynamics and how they interact with the other vortices of a BEC, determine the circulations of all vortices, and measure corresponding energy spectra.  This is an exceptionally challenging task!  But there are likely to be realistic methods for approaching at least some of these goals.  In 2D QT, methods for in situ imaging of vortices in the plane of a highly oblate BEC are currently being explored, and these show promise for use in multiple imaging methods that will allow for real-time probes of 2D QT vortex dynamics.  The 3D case is much more demanding.  

Atom interferometry presents an available method for circulation measurement \cite{Ino2001.PRL87.080402,Mat1999.PRL83.2498,Che2001.PRA64.031601}, and in principle may be used even in three dimensions.  In practice, whether in two or three dimensions, the basic application of atom interferometry in these cases would involve the interference of a turbulent BEC with a reference BEC. The reference BEC could be obtained either from coherently splitting a vortex-free condensate prior to driving one of the components into a turbulent state, or by creating two fully independent BECs of the same species. Trapping two BECs simultaneously is challenging; driving one and not the other into a turbulent state adds another layer of challenge that may be achievable with laser stirring beams, for example.  Matter wave interference that is suitable for resolving circulations of a high-vorticity BEC is another open experimental challenge.  Add to this the vortex tangles of 3D turbulence, and the experimental challenges are formidable, but not out of the question.  

Another experimental goal involves the generation of QT with minimal acoustic excitation so that forced QT can be reliably studied.  Forcing techniques are currently one of the main topics of numerical investigation.  As mentioned, a major experimental advance will occur with techniques that permit measurement of the kinetic energy spectrum of a BEC, although the correspondence of the total kinetic energy spectrum with the incompressible component is an open question.  Finally, on-demand vortex generation and manipulation techniques will allow for studies of vortex dynamics and interactions, as mentioned earlier.  

\section{Conclusions}

To summarise, atomic BECs are a highly tuneable system that hold much promise for the development of theoretical and experimental insights into some of the unanswered questions surrounding the theory of quantum turbulence. This holds particularly true regarding aspects of compressibility and QT, and the nature of 2D QT, both of which have not been explored prior to recent work with BECs.  With BECs, emerging methods for  controlled vortex creation may soon allow deterministic preparation of initial states necessary for investigations into the chaotic dynamics of few-vortex systems.  Routine creation of  highly oblate condensates provides the first system in which 2D quantum turbulence can be experimentally explored.  Atomic gas superfluids also provide numerous other opportunities not mentioned in this paper, such as possibilities to investigate QT in spinor systems, or in degenerate Fermi gases.  

Atomic BECs are also currently the most accessible system in which to study the small scale properties of turbulent vortex flow in three dimensions.  Theoretical investigations have begun to build up a picture of vortex dynamics and the processes contributing to the forcing and decay of turbulence at small scales and there is great scope for experimental verification in atomic BECs.  At large scales, features of classical turbulence is an emergent feature of quantum turbulence, suggesting that research with BEC systems may provide insight into some of the outstanding questions of turbulence.

\begin{acknowledgments}
BPA acknowledges funding from the US National Science Foundation, grant PHY-1205713. ACW acknowledges funding from EPSRC grant No. EP/H027777/1.
\end{acknowledgments}

\end{document}